# Compact, widely tunable ultrashort burst pulse generator using four mirrors

Rikako Tanaka[1,†], Keitaro Shimada[1,†], Ayumu Ishijima[2,*], Etsuko Kobayashi[1],
Hideharu Mikami[2], Ichiro Sakuma[1,3], and Keiichi Nakagawa[1]

[1]*Graduate School of Engineering, The University of Tokyo, Tokyo 113-8656, Japan*
[2]*Research Institute for Electronic Science, Hokkaido University, Hokkaido, 001-0021, Japan*
[3]*Medical Device Development and Regulation Research Center, The University of Tokyo, Tokyo 113-8656, Japan*
[†]*These authors contributed equally to this work.*
*ishijima@es.hokudai.ac.jp

**Ultrashort burst laser pulses serve as powerful tools for precise laser processing, broadband ultrafast spectroscopy, and high-speed laser-scanning microscopy. However, the performance of conventional burst pulse generators is limited by constraints in the pulse time interval variability, pulse energy variability, pulse number variability, and overall system complexity. Here, we present a compact burst pulse generator that offers a broad tuning range for pulse time intervals, along with control over the number of pulses and pulse energies within the burst. It consists of four mirrors, two of which are parallel to each other, and outputs pulses that are equally spaced both temporally and spatially. We demonstrated the generation of a burst laser pulse by shaping a single ultrashort laser pulse into six pulses with time intervals ranging from femtoseconds to nanoseconds. The pulse time intervals and energies were consistent with theoretical results.**

Recent advancements in high-power femtosecond (fs) laser sources and pulse-shaping technologies have enabled the generation of high-energy burst laser pulses—packets of high-repetition-rate pulses repeated at lower frequencies—spanning applications in broad research areas and industrial fields [1,2]. For example, in laser processing, the usage of bursts of femtosecond (fs) laser pulses at gigahertz (GHz) frequencies enhances the processing precision and material ejection efficiency compared with single-pulse laser processing [3,4]. In ultrafast spectroscopy, frequency-dependent material responses in the GHz to terahertz (THz) range can be probed using burst laser pulses with tunable temporal spacing [5–7]. Moreover, the use of burst laser pulses increases the data acquisition speed in laser-scanning microscopy, where each pulse is directed towards separate regions of a biological sample for the spatiotemporal multiplexing measurement of fluorescence [8–12]. These successes in various applications of burst laser pulses have led to a growing demand for further technological developments of the method to precisely control the time interval, energy, and number of pulses within a burst.

However, conventional burst pulse generators often lack one or more of the following attributes: wide tunability of the pulse time interval, variability in the energy for each pulse within the burst, variability in the pulse number, and system compactness. The generation of burst laser pulses can be classified into two methods: (1) adjusting the repetition frequency of the seed laser pulse using a pulse picker [e.g., electro-optic modulator (EOM) and acousto-optic modulator (AOM)] before entering the laser amplifier [13,14], and (2) shaping a single amplified fs laser pulse into sub-pulses using a free-space delay-line [15–20]. While the former method is widely used in industrial applications such as laser manufacturing, unwanted pulse distortion and optical damage to the gain medium limit control over the pulse time interval, pulse energy, and pulse number. In methods to generate burst laser pulses using a free-space delay-line such as Fabry-Perot cavities [15,16], spectrum circuit [17,18], spectrum shuttle [19], free-space angular-chirp-enhanced delay [20], and Deathstar [21–23], the pulse time interval corresponds to the round-trip length of circulation or reflection. Hence, its time interval can be easily changed by controlling the position of the optics. However, previous methods require complicated and large optical systems to achieve pulse time intervals of ~10-ps or less, posing challenges for system integration. Additionally, large optical systems inherently increase the optical path length, which reduces their robustness to environmental disturbances such as vibrations and air turbulence.

In this letter, we present a dual parallel reflection (DPR) burst pulse generator that provides a wide tuning range for the pulse time interval while allowing for control of the pulse energy and pulse number. It consists of four mirrors, two of which are parallel to each other, and outputs pulses that are equally spaced both temporally and spatially (Fig. 1). The first set of mirrors consists of a reflective mirror (mirror 1 in Fig. 1) and a partially reflective mirror (mirror 2 in Fig. 1), each with a rotation angle around the $y$-axis. The second set of mirrors consists of two reflective mirrors (mirrors 3 and 4 in Fig. 1), each with a rotation angle around the $x$ and $y$-axes. After passing through the first set of mirrors, the pulses are aligned parallel to the horizontal direction and pass through the upper part of mirror 3. Thereafter, the pulses are reflected by mirror 4 and undergo multiple reflections in the second set of mirrors. The pulses incident on the second set of mirrors are reflected different number of times, resulting in a final output of pulses that are equally aligned in the vertical direction. Consequently, the incident single pulses are repeatedly reflected and transmitted in the parallel set of mirrors, producing temporally and spatially equally-spaced pulses.

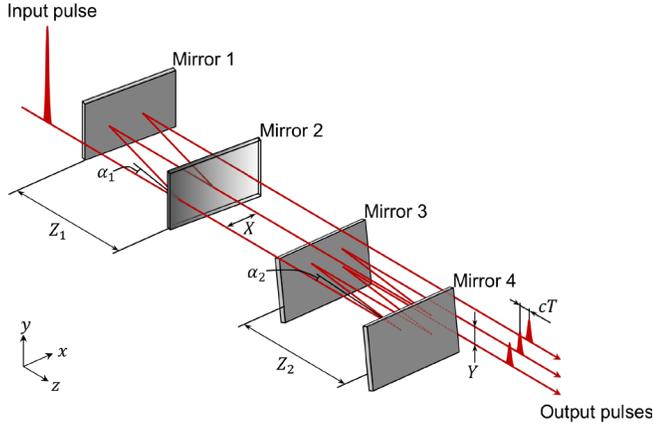

**Fig. 1.** Schematic of a DPR burst pulse generator with three output pulses. Mirrors 1 and 2, and mirrors 3 and 4 are parallel to each other. The dashed lines indicate rays obscured by mirror 4.

We can widely tune the pulse time interval by changing the distance between the second set of mirrors. The time interval of the pulse $T$ is expressed as

$$T = \frac{\left(Z_2 + \sqrt{Z_2^2 - X^2 - Y^2}\right) - \left(Z_1 + \sqrt{Z_1^2 - X^2}\right)}{c} \quad (1)$$

where $c$ is the speed of light, $X$ is the spatial distance between adjacent pulses after passing through the first set of mirrors, $Y$ is the spatial distance between adjacent pulses after passing through the second set of mirrors, $Z_1$ is the distance between the first set of mirrors in the $z$-direction $[Z_1 > (2X^2 + Y^2)/(2\sqrt{X^2 + Y^2})]$, and $Z_2$ is the distance between the second set of mirrors in the $z$-direction ($Z_2^2 > X^2 + Y^2$). Eq. 1 (see Supplement 1 for the derivation of the equation) is valid for $0 < \alpha_1 \leq \pi/4$ and $0 < \alpha_2 \leq \pi/4$, where $\alpha_1$ and $\alpha_2$ are the angle of incidence of the pulses on the first and second set of mirrors, respectively. The time delay produced by the first set of mirrors is canceled by the second set of mirrors at $Z_2 = Z_2'$, where $Z_2'$ is expressed as:

$$Z_2' = \frac{X^2 + Y^2 + \left(Z_1 + \sqrt{Z_1^2 - X^2}\right)^2}{2\left(Z_1 + \sqrt{Z_1^2 - X^2}\right)} \quad (2)$$

There is a linear relationship between the pulse time interval and the distance between the second set of mirrors in the $z$-direction when $Z_2^2 \gg X^2 + Y^2$ (Fig. 2). Moreover, without using the second set of mirrors, it is difficult to output a burst laser pulse with an fs-order pulse time interval for an input beam diameter larger than a few millimeters and an output number of pulses greater than three (see Supplement 1, Fig. S2).

The energy of each pulse within the burst can be controlled by changing the reflectance and transmittance of the partially reflective mirror (mirror 2 in Fig. 1). The transmittance profile of a partially reflective mirror is expressed as

$$T_i = \frac{I_i}{(1-r')I_0 r^{2M-i-1} \prod_{k=1}^{i-1} R_k} \quad (3)$$

where $i$ is an integer denoting the pulse number ($i = 1$ denotes the pulse transmitted first through the partially reflective mirror); $I_i$ is the pulse energy after passing through the second mirror; $r'$ is the reflectance on the back side of the partially reflective mirror (mirror 2 in Fig. 1); $r$ is the reflectance of the reflective mirror (mirrors 1, 3, and 4 in Fig. 1); $M$ is the total number of output pulses; and $R_i$ is the reflectance of the partially reflective mirror in the region traversed by the $i$-th pulse (see Supplement 1 for the derivation of the equation). By designing a partially reflective mirror with the transmittance profile based on Eq. 3, the desired energy for each pulse within the burst is obtained. Fig. 3(a) shows the pulse energy of the burst laser pulse for output pulses of twenty ($M = 20$). For the calculations, the energy losses from the DPR burst pulse generator were neglected ($r = 1$, $r' = 0$, and $I_0 = \sum_{k=1}^{M} I_k$). Three different transmittance profiles of the partially reflective mirror were assumed in the calculations [Fig. 3(b)]. The DPR burst pulse generator enables tailoring burst laser pulses with various temporal patterns by designing the transmittance profile of a partially reflective mirror. Noted that as shown in Fig. 3(b), it is impossible to completely match the total energy of the burst laser pulse to the pulse energy entering the DPR burst pulse generator ($I_0 > \sum_{k=1}^{M} I_k$) because finite sections of the Gaussian distribution are discretized to produce a Gaussian-distributed burst pulse. Therefore, unlike the other transmittance profiles of the partially reflecting mirrors, the transmittance in the area through which the 20th pulse passed did not reach 100%.

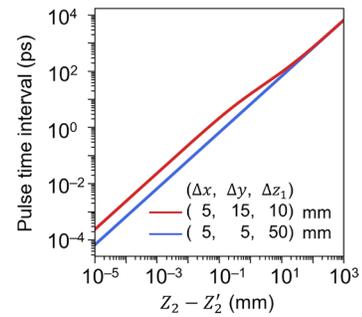

**Fig. 2.** Theoretical plot of the pulse time interval with the distance between the second set of mirrors in the $z$-direction. The offset on the distance between the second set of mirrors ($Z_2'$) is subtracted on the horizontal axis.

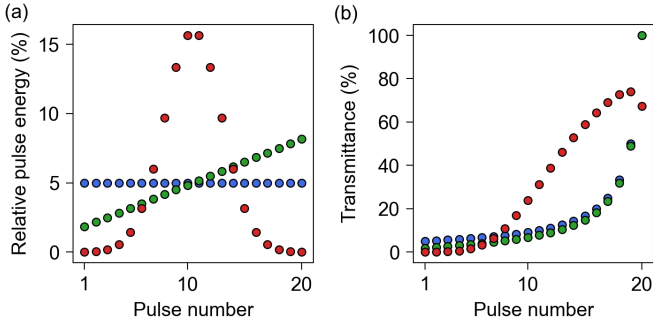

**Fig. 3.** Theoretical plots of the pulse energy for each pulse within the burst. (a) Output pulse energy distribution for a pulse number of twenty (red: Gaussian, green: linear increase, blue: constant). (b) Transmittance profile of a partially reflective mirror to produce the output pulse energy distribution shown in (a). The correspondence between the output pulse energy distribution and transmittance profile is shown in the same color plots.

To evaluate the performance of the DPR burst pulse generator, we constructed it in a 100 mm × 500 mm space. As a proof-of-principle experimental setup, we used a step-density filter (32-599, Edmund) as a partially reflective mirror to produce six pulses. The distance between the first set of mirrors was set to $Z_1 = 46.8$ mm and the spatial distance between adjacent pulses after passing through the first set of mirrors was set to $X = 4.6$ mm. The spatial distance between the adjacent pulses after passing through the second set of mirrors was set to $Y = 3.2$ mm. A mode-locked Ti:sapphire laser with a chirped pulse amplifier system (Astrella-USP-1K, Coherent) was used as the light source to generate an ultrashort pulse laser with a central wavelength of 803 nm, bandwidth of 35 nm, and pulse width of 35 fs at a repetition rate of 1 kHz.

We first demonstrated the generation of burst laser pulses with pulse time intervals ranging from femtoseconds to nanoseconds. The pulse time interval was changed by adjusting the distance between the second set of mirrors in the $z$-direction, while maintaining the spatial distance between adjacent pulses. The time intervals of pulses exceeding sub-nanoseconds were measured using a 33-GHz photodiode (DPO7OE1, Tektronix) with a 23-GHz oscilloscope (MSO72304DX, Tektronix). The time intervals of pulses of picoseconds or less were measured using autocorrelation. For the cross-correlation measurements, the pulse was split using a beam splitter before entering the DPR burst pulse generator to create a reference pulse. A delay line was placed in the reference beam path. The burst laser pulse and reference pulse were focused onto a beta barium borate crystal (5 mm ×5 mm ×2 mm) using a lens. A frequency-doubled laser pulse was detected using a photodiode (DET36A/M, Thorlabs) connected to an oscilloscope. The measured pulse intervals agreed with the theoretical curves calculated using Eq. 1 [Fig. 4(a)].

Next, the energy of each pulse within a burst was measured. The theoretical values were calculated by substituting the transmittance and reflectance of the partially reflective mirror into Eq. 3. The pulse energy was measured using an energy meter (PM100D, Thorlabs). As shown in Fig. 4(b), the measured energy of the pulses reflects the transmittance profile of the partially reflective mirror. These results indicate that the energy of each pulse within the burst can be controlled by the transmittance profile of the partially reflective mirror with negligible laser power loss.

We next compare the DPR burst pulse generator with previous burst pulse generators. First, the DPR burst pulse generator uses fewer optical elements and requires a shorter optical path length compared to Deathstar [21–23], which has a similar tunable range of the pulse time interval, pulse energy, and pulse number within the burst (see Supplement 1). Therefore, our method is potentially robust to environmental disturbances such as vibrations and air turbulence. However, the DPR burst pulse generator requires an adjustment of the mirror angle in addition to changing the spacing of the second set of mirrors to change the pulse time interval. Second, the DPR burst pulse generator outputs pulses that are equally spaced both temporally and spatially, whereas other burst generators such as the reverberation loop [9], Fabry-Perot cavities [15,16], spectrum circuits [17,18], and spectrum shuttle [19], output collinear burst laser pulses. Therefore, the maximum aperture of the objective lens cannot be used to focus the burst laser pulses on the sample. Although the DPR burst pulse generator can output collinear burst laser pulses by placing a retro-reflecting mirror immediately after the second set of mirrors, such an approach results in a power loss.

The DPR burst pulse generator is expected to have a wide range of scientific and industrial applications. Because it can generate burst laser pulses at fs time intervals, it can be used as a burst generator for single-shot ultrafast spectroscopy [24,25] and THz burst-mode laser processing [26,27]. Furthermore, the use of burst laser pulses to generate multi-cycle longitudinal waves can potentially enhance the acoustic signal from biological samples and increase the signal-to-noise ratio [5,21]. Therefore, the ability to generate burst laser pulses on sub-nanosecond timescales enables the possibility of using the DPR burst pulse generator in phonon spectroscopy to study cell biomechanics, where the Brillouin frequency of biological cells is typically on the order of ~1 GHz [28–30].

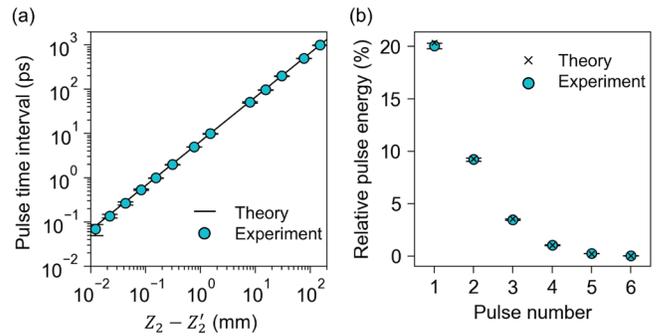

**Fig. 4.** Producing GHz to THz burst laser pulses by a DPR burst pulse generator. (a) The pulse time intervals ranging from 70±20 fs to 1.000±0.003 ns adjusted by the distance between the second set of mirrors in the $z$-direction. The error bar represents the standard deviation of the pulse intervals among six pulses. (b) The energy for each pulse within the burst. The error bar represents the standard deviation of the energy for 60,000 pulses.

In conclusion, we presented an ultrashort burst pulse generator that provides a wide tunable range of pulse time intervals while allowing control of the pulse number and pulse energy. The generation of burst laser pulses was demonstrated by shaping a single femtosecond laser pulse into six pulses with pulse intervals ranging from femtoseconds to nanoseconds, closely matching their pulse time intervals and energies with theoretical results. Our simple and compact burst pulse generator has the potential to replace conventional non-collinear burst pulse generators, enabling precise control of the pulse time interval, pulse number, and pulse energy.

**Funding.** JST PRESTO (JPMJPR1902), Japan Society for the Promotion of Science (23K18584), Basic Research Program of the Network Joint Research Center for Materials and Devices, Murata Science Foundation, and Sumitomo Foundation.

**Acknowledgment.** We thank Takao Saiki for fruitful discussions.

**Disclosures.** The authors declare no conflicts of interest.

**Data Availability.** Data underlying the results presented in this paper are available from the corresponding author upon reasonable request.

**Supplement Document.** See Supplement 1 for supporting content.

# COMPACT, WIDELY TUNABLE ULTRASHORT BURST PULSE GENERATOR USING FOUR MIRRORS: SUPPLEMENTAL DOCUMENT

## 1. Derivation of the pulse time interval

The dual parallel reflection (DPR) burst pulse generator can widely tune the pulse time interval by changing the distance between the second set of mirrors. The time interval of pulses $T$ is expressed as

$$T = (L_2 - L_1)/c \tag{S1}$$

where $L_1$ and $L_2$ ($L_2 > L_1$) are the differences in the optical path length between adjacent pulses in the first and second set of mirrors, respectively, and $c$ is the speed of light. First, $L_1$ can be expressed as

$$L_1 = |\boldsymbol{v_{1A}}| + \boldsymbol{v_{1A}} \cdot (0,0,-1) \tag{S2}$$

where $\boldsymbol{v_{1A}}$ is the vector representing the path of light from mirror 2 to mirror 1 (Fig. S1). The vector $\boldsymbol{v_{1A}}$ is expressed as

$$\boldsymbol{v_{1A}} = (\Delta x_{1A}, 0, \Delta z_{1A}) \tag{S3}$$

where $\Delta x_{1A}$ and $\Delta z_{1A}$ are the displacements along the $x$-axis and $z$-axis, respectively. The vector $\boldsymbol{v_{1B}}$, representing the path of light from mirror 1 to mirror 2 is expressed as

$$\boldsymbol{v_{1B}} = (0,0,\Delta z_{1B}) = (0,0,Z_1) \tag{S4}$$

where $\Delta z_{1B}$ is the displacement along the $z$-axis corresponding to the distance between the mirrors in the $z$-direction $Z_1$. As mirror 1 and 2 are parallel to each other, $Z_1$ can be expressed as

$$Z_1 = |\boldsymbol{v_{1B}}| = |\boldsymbol{v_{1A}}| = \sqrt{\Delta x_{1A}^2 + \Delta z_{1A}^2} \tag{S5}$$

Therefore, Eq. S2 can be re-written as

$$L_1 = Z_1 - \Delta z_{1A} = \sqrt{\Delta x_{1A}^2 + \Delta z_{1A}^2} - \Delta z_{1A} \tag{S6}$$

$\Delta z_{1A}$ can be expressed using the angle of incidence to the mirrors $\alpha_1$:

$$\Delta z_{1A} = -Z_1 \cos 2\alpha_1 \tag{S7}$$

$\cos 2\alpha_1$ can be expressed as

$$\cos 2\alpha_1 = \begin{cases} \dfrac{|\Delta z_{1A}|}{Z_1} & \left(0 < \alpha_1 \leq \dfrac{\pi}{4}\right) \\ -\dfrac{|\Delta z_{1A}|}{Z_1} & \left(\dfrac{\pi}{4} < \alpha_1 < \dfrac{\pi}{2}\right) \end{cases}$$

$$= \begin{cases} \dfrac{\sqrt{Z_1^2 - \Delta x_{1A}^2}}{Z_1} & \left(0 < \alpha_1 \leq \dfrac{\pi}{4}\right) \\ -\dfrac{\sqrt{Z_1^2 - \Delta x_{1A}^2}}{Z_1} & \left(\dfrac{\pi}{4} < \alpha_1 < \dfrac{\pi}{2}\right) \end{cases} \tag{S8}$$

In addition, $|\Delta x_{1A}|$ is equal to the spatial distance between adjacent pulses after passing through the first set of mirrors $X$. Therefore, Eq. S6 can be re-written as

$$L_1 = \begin{cases} Z_1 + \sqrt{Z_1^2 - X^2} & \left(0 < \alpha_1 \leq \dfrac{\pi}{4}\right) \\ Z_1 - \sqrt{Z_1^2 - X^2} & \left(\dfrac{\pi}{4} < \alpha_1 < \dfrac{\pi}{2}\right) \end{cases} \tag{S9}$$

The difference in the optical path length generated by the second set of mirrors is expressed as

$$L_2 = |\boldsymbol{v_{2A}}| + \boldsymbol{v_{2A}} \cdot (0,0,-1) \tag{S10}$$

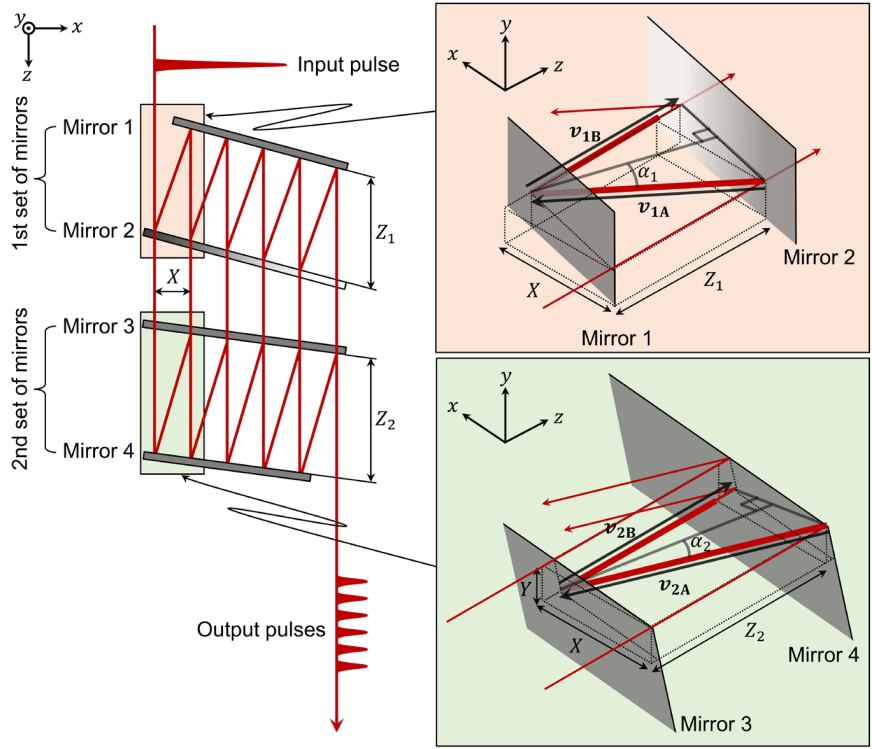

**Fig. S1.** Schematic of a DPR burst pulse generator. The orange and green insets show the pulse traveling between the first and second set of mirrors, respectively. The red bold line indicates the optical path length difference between adjacent pulses.

where $v_{2A}$ is a vector representing the path of light from mirror 4 to mirror 3. Vector $v_{2A}$ is expressed as

$$v_{2A} = (\Delta x_{2A}, \Delta y_{2A}, \Delta z_{2A}) \tag{S11}$$

where $\Delta x_{2A}$, $\Delta y_{2A}$, and $\Delta z_{2A}$ are the displacements along the x-axis, y-axis and z-axis, respectively. The vector $v_{2B}$, representing the path of light traveling from mirror 3 to mirror 4 is expressed as

$$v_{2B} = (0, 0, \Delta z_{2B}) = (0, 0, Z_2) \tag{S12}$$

where $\Delta z_{2B}$ is the displacement along the z-axis corresponding to the distance between the mirrors in the z-direction $Z_2$. As mirrors 3 and 4 are parallel to each other, $Z_2$ can be expressed as

$$Z_2 = |v_{2B}| = |v_{2A}| = \sqrt{\Delta x_{2A}^2 + \Delta y_{2A}^2 + \Delta z_{2A}^2} \tag{S13}$$

Therefore, Eq. S10 can be re-written as

$$L_2 = Z_2 - \Delta z_{2A} = \sqrt{\Delta x_{2A}^2 + \Delta y_{2A}^2 + \Delta z_{2A}^2} - \Delta z_{2A} \tag{S14}$$

$\Delta z_{2A}$ can be expressed using the angle of incidence to the mirrors $\alpha_2$ as

$$\Delta z_{2A} = -Z_2 \cos 2\alpha_2 \tag{S15}$$

$\cos 2\alpha_2$ can be expressed as

$$\cos 2\alpha_2 = \begin{cases} \dfrac{|\Delta z_{2A}|}{Z_2} & \left(0 < \alpha_2 \leq \dfrac{\pi}{4}\right) \\ -\dfrac{|\Delta z_{2A}|}{Z_2} & \left(\dfrac{\pi}{4} < \alpha_2 < \dfrac{\pi}{2}\right) \end{cases}$$

$$= \begin{cases} \dfrac{\sqrt{Z_2{}^2 - \Delta x_{2A}{}^2 - \Delta y_{2A}{}^2}}{Z_2} & \left(0 < \alpha_2 \leq \dfrac{\pi}{4}\right) \\ -\dfrac{\sqrt{Z_2{}^2 - \Delta x_{2A}{}^2 - \Delta y_{2A}{}^2}}{Z_2} & \left(\dfrac{\pi}{4} < \alpha_2 < \dfrac{\pi}{2}\right) \end{cases} \quad \text{(S16)}$$

In addition, $|\Delta x_{2A}|$ is equal to $X$, and $|\Delta y_{2A}|$ is equal to the spatial distance between adjacent pulses after passing through the second set of mirrors $Y$. Therefore, Eq. S14 can be re-written as

$$L_2 = \begin{cases} Z_2 + \sqrt{Z_2{}^2 - X^2 - Y^2} & \left(0 < \alpha_2 \leq \dfrac{\pi}{4}\right) \\ Z_2 - \sqrt{Z_2{}^2 - X^2 - Y^2} & \left(\dfrac{\pi}{4} < \alpha_2 < \dfrac{\pi}{2}\right) \end{cases} \quad \text{(S17)}$$

From Eqs. S1, S6 and S14, the pulse time interval $T$ is expressed as

$$T = \frac{\left(\sqrt{\Delta x_{2A}{}^2 + \Delta y_{2A}{}^2 + \Delta z_{2A}{}^2} - \Delta z_{2A}\right) - \left(\sqrt{\Delta x_{1A}{}^2 + \Delta z_{1A}{}^2} - \Delta z_{1A}\right)}{c} \quad \text{(S18)}$$

The time delay produced by the first set of mirrors is canceled by the second set of mirrors at $\Delta z_{2A} = \Delta z'_{2A}$, where $\Delta z'_{2A}$ is expressed as

$$\Delta z'_{2A} = \frac{\Delta x_{2A}{}^2 + \Delta y_{2A}{}^2 - \left(\sqrt{\Delta x_{1A}{}^2 + \Delta z_{1A}{}^2} - \Delta z_{1A}\right)^2}{2\left(\sqrt{\Delta x_{1A}{}^2 + \Delta z_{1A}{}^2} - \Delta z_{1A}\right)} \quad \text{(S19)}$$

In this letter, we address the conditions under which the angles of incidence $\alpha_1$ and $\alpha_2$ for the first and second set of mirrors, respectively, satisfy $0 < \alpha_1 \leq \pi/4$, $0 < \alpha_2 \leq \pi/4$. Under these conditions, the pulse time interval $T$ can be derived from Eqs. S1, S9, and S17 as follows:

$$T = \frac{\left(Z_2 + \sqrt{Z_2{}^2 - X^2 - Y^2}\right) - \left(Z_1 + \sqrt{Z_1{}^2 - X^2}\right)}{c} \quad \text{(S20)}$$

When $X$, $Y$ and $Z_1$ are fixed and $Z_2{}^2 \gg X^2 + Y^2$, $T$ is linearly related to $Z_2$

$$T \approx \frac{2}{c}Z_2 - \frac{Z_1 + \sqrt{Z_1{}^2 - X^2}}{c} \quad \text{(S21)}$$

When satisfying $0 < \alpha_1 \leq \pi/4, 0 < \alpha_2 \leq \pi/4$, the time delay produced by the first set of mirrors is canceled by the second set of mirrors at $Z_2 = Z'_2$, where $Z'_2$ is expressed as:

$$Z'_2 = \frac{X^2 + Y^2 + \left(Z_1 + \sqrt{Z_1{}^2 - X^2}\right)^2}{2\left(Z_1 + \sqrt{Z_1{}^2 - X^2}\right)} \quad \text{(S22)}$$

For $T = 0$ to hold within the range $0 < \alpha_2 \leq \pi/4$, it is necessary to satisfy $\Delta z'_{2A} < 0$. Based on Eqs. S5 and S19, this condition can be expressed as

$$Z_1 > \frac{2X^2 + Y^2}{2\sqrt{X^2 + Y^2}} \quad \text{(S23)}$$

## 2. Derivation of pulse energy for each pulse within the burst

The energy of each pulse within the burst can be controlled by changing the reflectance and transmittance of the partially reflective mirror (mirror 2 in Fig. S1). The pulse energy after passing through the first mirror set $I'_i$ is calculated as

$$I'_i = T_i(1 - r')I_0 r^{i-1} \prod_{k=1}^{i-1} R_k \qquad (S24)$$

where $i$ is an integer denoting the pulse number ($i = 1$ denotes the pulse transmitted first through the partially reflective mirror), $I_0$ is the pulse energy entering the DPR burst pulse generator, $R_i$ and $T_i$ are the reflectance and transmittance of the partially reflective mirror in the region traversed by the $i$th pulse, respectively, $r'$ is the reflectance on the back side of the partially reflective mirror, and $r$ is the reflectance of the reflective mirror (mirror 1, 2, and 3 in Fig. S1). The pulse energy after passing through the second set of mirrors $I_i$ is calculated as

$$I_i = I'_i \times r^{2(M-i)} \qquad (S25)$$

where $M$ is the total number of pulses. From Eq. S24 and S25, the transmittance profile of the partial reflective mirrors is expressed as

$$T_i = \frac{I_i}{(1 - r')I_0 r^{2M-i-1} \prod_{k=1}^{i-1} R_k} \qquad (S26)$$

By designing a partial reflective mirror with transmittance profile based on Eq. S26, the desired pulse energy for each pulse within the burst can be obtained.

## 3. Minimum pulse time interval produced by first set of mirrors

Burst pulses can be generated using only the first set of mirrors. This configuration is referred to as single parallel reflection (SPR). The difference between burst pulse generation using SPR and DPR burst pulse generator is the range of feasible pulse time intervals. From Eqs. S6 and S7, the pulse time interval of the SPR burst pulse generator can be expressed as

$$T = \frac{\sqrt{\Delta x_{1A}^2 + \Delta z_{1A}^2} - \Delta z_{1A}}{c} = \frac{Z_1(1 + \cos 2\alpha_1)}{c} = \frac{Z_1 \cos^2 \alpha_1}{c} \quad (S27)$$

The distance between the pulses along the mirrors is

$$l_{mirror} = 2Z_1 \sin \alpha_1 = \frac{X}{\cos \alpha_1} \quad (S28)$$

Therefore, Eq. S27 can be re-written as

$$T = \frac{l_{mirror} \cos^2 \alpha_1}{2c \sin \alpha_1} = \frac{l_{mirror} \frac{X^2}{l_{mirror}^2}}{2c \sqrt{1 - \frac{X^2}{l_{mirror}^2}}} = \frac{X^2}{c\sqrt{l_{mirror}^2 - X^2}} \quad (S29)$$

The condition that the pulse is not chipped by mirror 1 is

$$X \geq D \quad (S30)$$

where $D$ is the beam diameter. The length of the partially reflective mirror (mirror 2 in Fig. S1) $L_{mirror}$ is

$$L_{mirror} \geq l_{mirror}(M - 1) \quad (S31)$$

From Eqs. S30 and S31, Eq. S29 can be re-written as

$$T \geq \frac{D^2}{c\sqrt{\left(\frac{L_{mirror}}{M-1}\right)^2 - D^2}} \quad (S32)$$

Fig. S2 shows the minimum pulse time interval with beam diameters $D$ of 0.5, 1.0, 2.0 and 4.0-mm. It is difficult to achieve pulse time intervals of ~100 fs without using a second set of mirrors for a beam diameter of more than ~1 mm and more than three pulses.

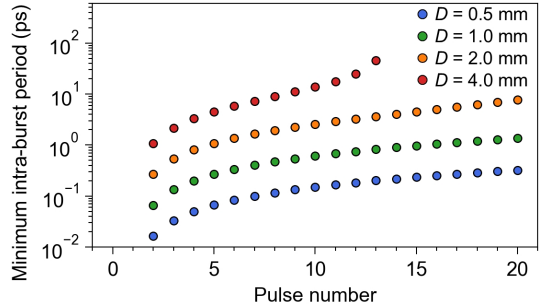

**Fig. S2.** Minimum pulse time interval produced by the first set of mirrors (SPR). The length of partial reflective mirror (mirror 2 in Fig. S1) $L_{mirror}$ was fixed to 50 mm.

## 4. Optical path length difference between DPR burst pulse generator and Deathstar

The beam fluctuates spatially because of environmental factors such as temperature changes and air flows. In general, these effects increase with propagation distance when the beam propagates freely. Therefore, the optical path length required to achieve the desired pulse time interval should be minimized. Here, we derived the optical path length difference of the pulse with the highest number of reflections in the second set of mirrors ($i = 1$) between the DPR burst pulse generator ($l_{\text{DPR}}$) and Deathstar ($l_{\text{Deathstar}}$). From Fig. S3, the optical path length difference is expressed as

$$l_{\text{Deathstar}} - l_{\text{DPR}} = 3(M-1)X + 2X - Z' \tag{S33}$$

where $Z' = (M-1)X \tan \alpha_1$. As $0 < \alpha_1 \leq \pi/4$, $l_{\text{Deathstar}} - l_{\text{DPR}}$ is within the range of

$$2MX < l_{\text{Deathstar}} - l_{\text{DPR}} \leq (3M-1)X \tag{S34}$$

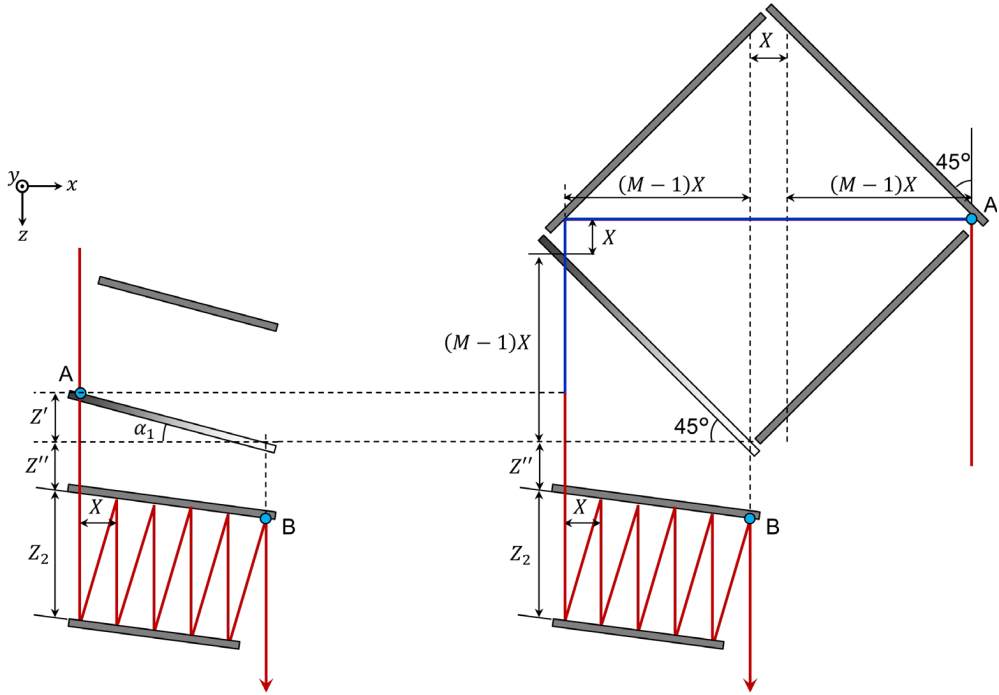

**Fig. S3.** Schematic of the optical path length difference between DPR burst pulse generator and Deathstar (blue line).